# Conceptual framework for performing simultaneous fold and sequence optimization in multi-scale protein modeling

István Kolossváry[1,2,*]

[1]*Department of Chemistry, Budapest University of Technology and Economics,*
*H-1111 Budapest, Hungary*

[2]*BIOKOL Research, LLC, Madison, New Jersey 07940, USA*

[*]Correspondence: Istvan@Kolossvary.hu

We present a dual optimization concept of predicting optimal sequences as well as optimal folds of off-lattice protein models in the context of multi-scale modeling. We validate the utility of the recently introduced hidden-force Monte Carlo optimization algorithm by finding significantly lower energy folds for minimalist and detailed protein models than previously reported. Further, we also find the protein sequence that yields the lowest energy fold amongst all sequences for a given chain length and residue mixture. In particular, for protein models with a binary sequence, we show that the sequence-optimized folds form more compact cores than the lowest energy folds of the historically fixed, Fibonacci-series sequences of chain lengths of 13, 21, 34, 55, and 89. We then extend our search algorithm to use UNRES, one of the leading united-residue protein force fields. Our combined fold and sequence optimization on three test proteins reveal an inherent bias in UNRES favoring alpha helical structures even when secondary structure prediction clearly suggests only beta sheets besides random coil, and virtually no helices. One test in particular, a triple-stranded antiparallel beta-sheet protein domain, demonstrates that by permutations of its sequence UNRES re-folds this structure into a perfect alpha helix but, in fact, the helix is just an artefact of the force field, the structure quickly unfolds in all-atom state-of-the-art molecular dynamics simulation.

I. INTRODUCTION

We recently introduced the hidden-force algorithm (HFA), a global Monte Carlo optimization method and used it to predict low-energy binary Lennard-Jones (BLJ) cluster configurations [1]. In this communication, we apply HFA to find the optimal fold of simple and detailed protein models. Further, we also find the protein sequence that yields the lowest energy fold amongst all sequences for a given chain length and residue mixture. In particular, we study protein AB (PAB) models [2, 3]. These have close similarity to BLJ models. Despite their minimalism, PAB models mimic a basic feature of protein folding—the formation of a hydrophobic core. Similar to BLJ models PAB models consist of only two types of residues, denoted A and B. The interaction energy between two A residues (AA interaction) is twice as strong as AB or BB interactions to promote core formation. The interaction potential has a pseudo Lennard-Jones form and the total potential energy includes angle-bending and torsion terms to account for local interactions. Historically, PAB models use a Fibonacci series sequence protein with chain lengths of 13, 21, 34, 55, and 89 [4, 5]. At first, we revisit the Fibonacci sequences and find that much lower energy folds exist than previously reported. Moreover, the new putative global minima show topological features qualitatively different from real proteins. The new low-energy folds are deeply knotted indicative of a flaw in the PAB model. Simplifying the local interaction terms introduced to extend the PAB model to three dimensions [3], we obtain new realistic low-energy folds without knots. Using this simplified potential we use HFA optimization to find non-Fibonacci sequences that fold into still lower energy structures with more compact cores comprised of A residues.

We then apply HFA optimization to six test proteins using UNRES, one of the leading united-residue force fields [6, 7, 8] that has been highly successful in the past fifteen years of CASP competition (Critical Assessment of protein Structure Prediction [9]) [10]. First, as a validation of HFA on this sophisticated two-bead model, we compare our global minima to those found by UNRES/CSA [11] and for five out of the six proteins find lower energy folds. We also find, however, that these new global minima all represent slightly unfolded structures, which is indicative of a flaw in the force field and/or a flaw in the concept of locating the global minimum of a potential energy function as a means to identify the global minimum free-energy structure/fold. Furthermore, we probe the UNRES force field by once again applying sequence optimization in order to find sequences that fold into still lower energy structures, and find that UNRES has a strong tendency to fold sequences into alpha helices. One test in particular, a WW domain (1E0L), which is the smallest monomeric triple-stranded antiparallel beta-sheet protein domain, demonstrates that by permutations of its sequence HFA/UNRES re-folds this structure into a perfect alpha helix.

## II. MODEL AND ALGORITHMIC DETAILS

PAB models are one of the minimalist protein models [12]. We used the model by Irbäck *et. al* [3], a well-studied three-dimensional extension of the original two-dimensional PAB model by Stillinger *et. al* [2]. In this model proteins consist of two types of residues, A and B. Each residue represents a Cα atom. The Cα–Cα bonds are set to unit length and the potential energy is:

$$E = -\kappa_1 \sum_{i=1}^{N-2} \cos C\alpha_{i,i+1,i+2} - \kappa_2 \sum_{i=1}^{N-3} \cos C\alpha_{i,i+1,i+2,i+3} + \sum_{i=1}^{N-2} \sum_{j=i+2}^{N} 4\varepsilon(\sigma_i, \sigma_j)\left(\frac{1}{r_{ij}^{12}} - \frac{1}{r_{ij}^{6}}\right).$$

Eq. 1

The first two sums represent local interactions as angle-bending or torsion energies, involving three or four consecutive Cα atoms, respectively. The last double sum is a pseudo Lennard-Jones (LJ) potential representing long-range interactions. $N$ is the number of residues, $r_{ij}$ is the distance between two residues, $\varepsilon(\sigma_i, \sigma_j)$ is the residue pair-specific LJ well-minimum depth, and $\kappa_1$ and $\kappa_2$ are empirical parameters determined by Monte Carlo simulations to reproduce qualitatively Cα angle and torsion distributions of proteins in the Protein Data Bank (PDB) [3]. We used $\kappa_1 = -1$ and $\kappa_2 = 0.5$. $\varepsilon(\sigma_i, \sigma_j)$ favor the formation of a core of A residues analogous to the hydrophobic core of real proteins: $\varepsilon(A, A) = 1$ and $\varepsilon(A, B) = \varepsilon(B, B) = 0.5$. LJ interactions between adjacent residues are excluded. In lieu of constraints we add strong harmonic distance terms to keep Cα–Cα bond lengths at unity and flat-bottom angle-bending terms to disallow near linear Cα–Cα–Cα bond angles (within two degrees). Linear bond angles cause a catastrophic physical divergence in the torsion term of the PAB model.

The UNRES model is a sophisticated two-bead representation of the polypeptide chain. Each amino acid residue is represented by two interaction centers, one is the halfway point between two consecutive Cα atoms (the Cα atoms themselves serve only as geometric reference points), and the other interaction center is the center of mass of the side chain that is modeled as an ellipsoid with two rotational degrees of freedom with respect to the backbone. Backbone flexibility itself is allowed by varying virtual-bond angles and virtual-bond dihedral angles along three or four Cα atoms, respectively. The UNRES force field includes numerous bonded and non-bonded terms as well as implicit contributions from the interaction of the side chain with the solvent [6, 7, 8] and has been parameterized utilizing the so-called hierarchical design of the potential-energy landscape employing multiple training proteins simultaneously [13].

The hidden-force algorithm [1] exploits that, though the gradient components of an additive potential sum to zero at a local minimum, each component's magnitude is generally nonzero. Disrupting this network of opposing forces (negative of the gradient) can result in the collective rearrangement of cluster atoms. Using a tug-of-war analogy to describe the basic HFA move, some players (atoms) simultaneously drop their ropes (drop their contributions to the potential). The remaining players then rearrange due to their net nonzero tugging and reach a partial impasse. Then the dropouts resume tugging until a new total impasse is achieved. There is no guarantee that the resulting cluster configuration will be lower in energy than the starting configuration, but we found HFA to be an exceptionally successful move set in a Monte Carlo cluster minimization. HFA trial configurations are highly dependent on the starting configurations since the moves are driven by forces already present—making the HFA Monte Carlo search non-Markovian.

Algorithmically, PAB models can be treated similar to BLJ clusters with slightly different LJ terms and additional geometric constraints favoring protein like chains. We used the same algorithm and software described in detail in [1] with two notable differences: (1) when the basic HFA move is applied to a given local minimum-energy configuration, only the pseudo LJ terms are dropped; the remaining terms stay in effect to preserve the chain geometry. (2) Single or multiple mutations are utilized by swapping the types of randomly selected residues rather than flipping the type of a single residue at a time, in order to keep the composition fixed while varying the sequence. A BLJ cluster of fixed size is fully determined by its A/B composition, but PAB models also depend on the *sequence* of the residues. Therefore, while swapping A and B particles in a BLJ cluster will not change its identity, swapping A and B residues in a PAB protein will be a mutation. Further, in [1], we found the optimal A/B composition of a BLJ cluster of fixed size with lowest energy. For PAB models, the optimal A/B composition is trivial, a sequence of all A residues (AA interactions are stronger than AB or BB interactions). The more interesting and challenging mutation study we carry out keeps the A/B composition fixed and optimizes over sequences to find the lowest energy fold.

Employment of the UNRES model requires further significant but straightforward changes in the HFA algorithm with respect to how it is utilized with PAB models. (1) When the basic HFA move is applied, instead of simply dropping the non-bonded interactions, the selected residue is temporarily mutated into GLY, and it keeps all of its interactions. (2) Since UNRES is a two-bead model, every local minimization step is combined with side chain optimization along the full length of the polypeptide chain. Moreover, akin to the PAB studies we mutate the UNRES models by swapping one or more pairs of residues keeping the overall residue composition fixed and optimize over sequences to find the lowest energy fold. It should be noted, however, that while such residue swaps readily allow direct comparison of the energies of different sequence mutants in the minimal PAB model, this is not trivial with UNRES. Similar to all-atom molecular mechanics force fields, UNRES energy is not absolute and, therefore, energies of two different molecules are generally not comparable. Nonetheless, when comparing the energies of two different peptides with the same residue composition, but subject to single or multiple residue swaps and altogether having different residue sequences, is an exception. The energies of such peptides (we shall call them permutation isomers) are indeed comparable in any molecular mechanics force field including UNRES devoid of sequence dependent terms. UNRES does have the built-in capability of employing sequence dependent terms, but the current parameterization omits them.

## III. RESULTS AND DISCUSSION

To test HFA against existing optimization methods for PAB models, we run HFA searches on Fibonacci sequences in Table I and compare our putative global minimum-energy folds with those found in [4, 5]. Table I clearly demonstrates that HFA Monte Carlo search is efficient searching the fold space of PAB models. The difference in energy does not carry information about structural differences, though. Direct comparison was, unfortunately, not possible because the coordinates of the structures reported in [4, 5] were not published. Nonetheless, based on the visualizations in [4, 5], FIG. 1 clearly shows that our new minima belong to a different topological class indicative of a flaw in PAB models. With the exception of S_13, which is simply too short and S_21 that forms a simple trefoil knot (one end of the chain folding back through a loop), every other fold is deeply knotted (coordinates are listed in the Supplementary Material). Knotted protein structures occur naturally [14, 15], however, this level of knot formation cannot be found in the PDB [16]. With $\kappa_1 = -1$ and $\kappa_2 = 0.5$ [3], the simple cosine terms favor a 180 degree bond angle and a zero degree torsion angle. In real proteins, the bond angle distribution has a well-defined structure and bond angles strictly fall within the range of 85-145 degrees [3]. Torsion angles, on the other hand, are more uniform; there are no disallowed values. Nevertheless, zero degree torsion angles are rare. Eq. 1 represents an additive potential and even though the PAB models will always be frustrated in three dimensions, we can expect that at or close to the global minimum many individual energy terms will be close to their minimum values. This is exactly what we can see in FIG. 1; numerous bond angles are close to 180 degrees (kept away from exact linearity by the flat bottom angle-bending term) and numerous torsion angles are very close to zero degrees. This is the geometrical basis for forming knots in the ground sate, never seen in real proteins.

Table I. Energies of the new putative global minima listed in ε units, found by HFA for five Fibonacci sequences. The * operator in the Sequence column means concatenation. The previously reported energies were taken from [4, 5].

| Model identifier with chain length | Sequence | Lowest energy [ε] previously reported | Putative global minimum energy [ε] |
|---|---|---|---|
| S_13 | ABBABBABABBAB | -26.507 | -27.171 |
| S_21 | BABABBAB * S_13 | -52.934 | -56.409 |
| S_34 | S_13 * S_21 | -98.357 | -106.115 |
| S_55 | S_21 * S_34 | -176.691 | -190.579 |
| S_89 | S_34 * S_55 | -311.614 | -325.578 |

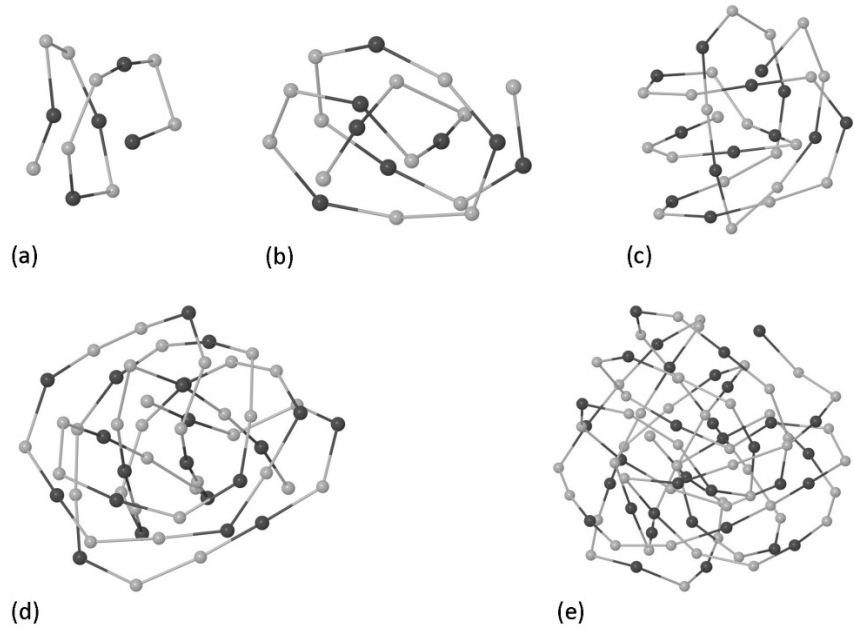

FIG. 1. Putative global minimum-energy folds found for 5 Fibonacci sequences using the potential in Eq. 1. Dark balls represent type A ("hydrophobic") residues and light grey balls are type B residues. Sequences and energies are given in Table I. (a) S_13, (b) S_21, (c) S_34, (d) S_55, and (e) S_89.

Ideally, more sophisticated potentials should be derived from the actual Cα bond angles and torsion angles found in the PDB via, e.g., the Boltzmann inversion method [12]. Keeping with the minimalist spirit of PAB models, however, we drop the angle-bending and torsion terms altogether, but we add a flat-bottom harmonic angle term that disallows Cα–Cα–Cα angles outside the 85-145 degree range. Using this potential we find that the Fibonacci sequences fold into globular structures with no tendency to form knots, and with a "hydrophobic" core. We then carried out sequence optimization using this potential. The left hand side of FIG. 2 shows the putative global minima found for the five Fibonacci sequences and the right hand side shows the lowest energy folds after sequence mutations were applied to the Fibonacci sequences (coordinates are available in the Supplementary Material). Visual inspection confirms the simplified PAB model yields compact folds with a clear core and no knots. The cores are more compact in the mutated sequences, which is quantified in Table II by the radius of gyration of the core (type A) residues. Table II also lists the energy drop after sequence optimization with respect to the putative global minimum energy of the Fibonacci sequence.

Table II. Fold optimization via sequence mutation. Energy drop and radius of gyration of the "hydrophobic" core formed by the type A residues. Energy is computed with the LJ-only potential (see text). The optimal sequences, coordinates, and absolute energies are listed in the Supplementary Material.

| Model identifier with chain length | Energy drop [$\varepsilon$] relative to Fibonacci sequence global min. | Radius of gyration of the core in Fibonacci sequence global min. | Radius of gyration of the core in optimal-sequence global min. |
|---|---|---|---|
| S_13 | -0.346 | 0.203 | 0.156 |
| S_21 | -1.369 | 0.288 | 0.223 |
| S_34 | -2.748 | 0.336 | 0.281 |
| S_55 | -9.074 | 0.371 | 0.327 |
| S_89 | -18.576 | 0.460 | 0.365 |

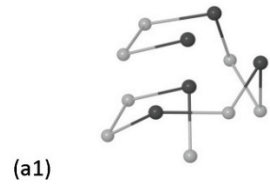 (a1)
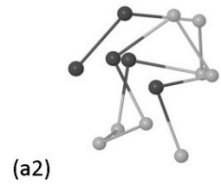 (a2)
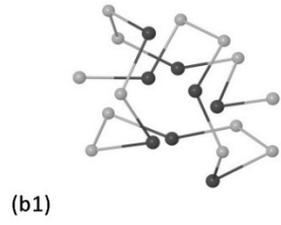 (b1)
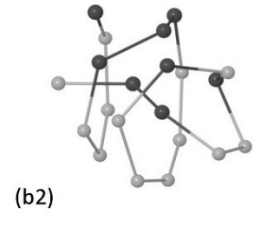 (b2)
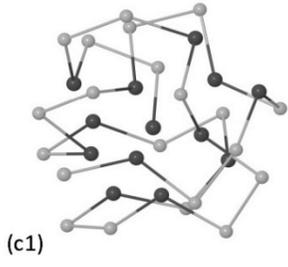 (c1)
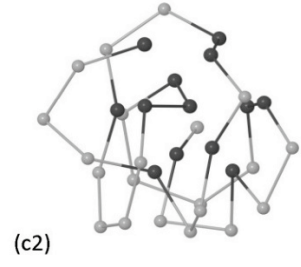 (c2)
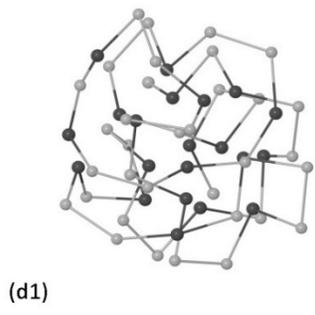 (d1)
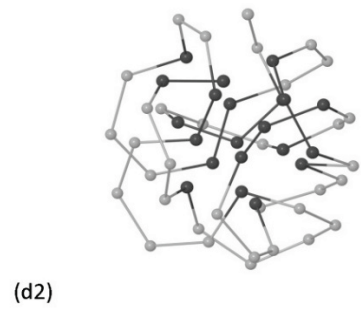 (d2)
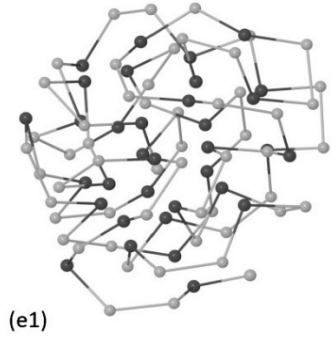 (e1)
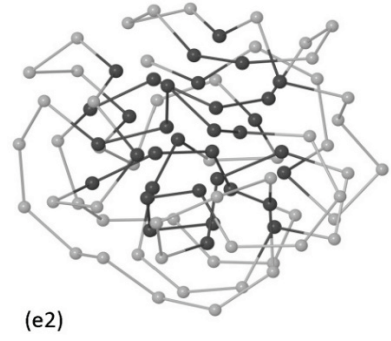 (e2)

FIG. 2. Putative global minimum-energy folds using the simplified LJ-only potential (see text). On the left hand side the Fibonacci sequences are shown and on the right hand side the optimal folds are displayed that were found after sequence optimization. Dark balls represent type A ("hydrophobic") residues and light grey balls are type B residues. The optimal sequences, coordinates, and energies are listed in the Supplementary Material. (a1, a2) S_13, (b1, b2) S_21, (c1, c2) S_34, (d1, d2) S_55, and (e1, e2) S_89.

To further test HFA against UNRES' native conformational space annealing (CSA) [11] global-optimization method, we run HFA searches on six test proteins (1BDD, 1GAB, 1LQ7, 1CLB, 1E0G, and 1IGD) that were essential in parameterizing UNRES. Among the force field options in UNRES we employ 4P (parameterized using four training proteins simultaneously) that is most adequate for small proteins of any fold including alpha, beta, and alpha/beta [13, 17]. Table III demonstrates that HFA Monte Carlo search is also efficient searching the fold space of sophisticated UNRES models. Similar to the PAB study, the difference in energy itself does not carry information about structural differences and direct comparison was, unfortunately, not possible because the coordinates of the structures reported in [18] were not published with the exception of 1IGD for which the complete CSA results are available in the UNRES download kit [19]. Nevertheless, the results point to noteworthy anomalies. Similar energies and similar RMS distances found with 1BDD, 1CLB, and 1E0G suggest that CSA and HFA searches located the same ground state folds for these chains. Moreover one could even argue that since the HFA ground state for 1GAB is significantly lower in energy than the lowest lying CSA structure and at the same time the HFA fold is 0.3 Å closer to the experimental structure in C-alpha RMS distance than the CSA minimum; lowering the energy might have the general effect of actually getting closer to the native structure, and thereby providing strong support for the force field. However, the two remaining proteins in Table III crush such hopes. HFA generated a fold for 1LQ7 that has virtually the same energy as the CSA ground state, yet the HFA fold is more than 4 Å farther away from the experimental structure than the quite accurate CSA minimum. 1IGD yields an even more negative result where the HFA ground state is, again, over 4 Å farther away than the CSA minimum, but the HFA fold has significantly (11 kcal/mol) lower energy. The C-alpha RMS distance between the CSA and HFA folds is 8.8 Å and visual inspection clearly shows that in the HFA ground state the alpha helix and both antiparallel beta strands start to unfold. The C-alpha coordinates of all six of the HFA putative global minimum-energy structures are listed in the Supplementary Material.

Table III. UNRES/4P energies of the lowest lying structures of six test proteins and their C-alpha RMS distances from the corresponding experimental structure found by CSA [18] and HFA, respectively. The number of residues are shown in parentheses. The C-alpha coordinates of the HFA putative global minimum-energy structures are listed in the Supplementary Material.

| PDB ID | CSA glob. min. energy [kcal/mol] | HFA glob. min. energy [kcal/mol] | CSA RMSD to exp. struct. [Å] | HFA RMSD to exp. struct. [Å] |
|---|---|---|---|---|
| 1BDD (46) | -597 | -601 | 5.5 | 5.6 |
| 1GAB (47) | -669 | -681 | 2.9 | 2.6 |
| 1LQ7 (67) | -937 | -937 | 2.3 | 6.6 |
| 1CLB (75) | -1053 | -1054 | 5.1 | 5.2 |
| 1E0G (48) | -632 | -634 | 4.1 | 4.3 |
| 1IGD (61) | -741 | -752 | 5.6 | 9.9 |

We conclude our UNRES tests with running HFA sequence optimization on three proteins to learn more about the force field. We choose a pure alpha protein 1BDD, a mixed alpha/beta protein 1IGD, and a pure beta mini protein 1E0L (28 residues), which is the smallest monomeric triple-stranded antiparallel beta-sheet protein domain. We want to see if sequence optimization can provide stable folds lower in energy than those of the native sequence/fold and whether we can draw any general conclusions about UNRES in this regard as we do with the modified PAB model above. Sequence optimization involves periodically swapping pairs of residues followed by HFA fold search and thereby exploring the sequence space of permutation isomers. As noted above, UNRES energy is not absolute but energies of permutation isomers are readily comparable.

FIG. 3, FIG. 4, and FIG. 5 give vivid visual insight and to put it bluntly; everything folds into a helix. The left hand side of the figures show the HFA ground state structure of the native sequence (rainbow colors from red N-terminus to purple C-terminus) superposed on the experimental structure shown in gray and the right hand side presents the result of sequence optimization. The figures also include the native sequence vs. optimized sequence and the associated energy drop. The evident qualitative statement that can be drawn from the figures is that sequence optimization re-folds all three native structures into helical structures with significant energy stabilization. Moreover, as shown by explicit tube representation, PRO residues are either pushed out the termini or break the alpha helices in the interior of the chain as generally seen in real protein structures, and GLY residues tend to shift to flexible coil segments. One would of course be tempted to draw physical conjectures based on the sequence optimization data, but further analysis makes it clear that folding into these predominantly alpha helical structures is, in fact, just a serious artefact of the UNRES force field. We provide two types of evidence.

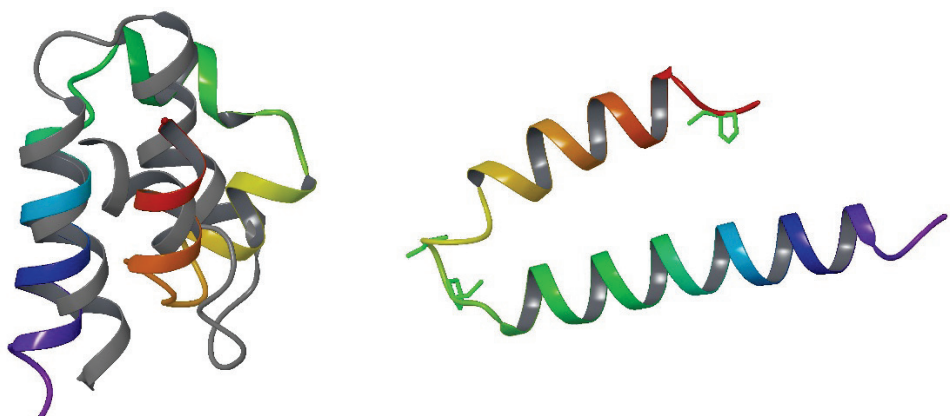

FIG. 3. Ground state sequence-permutation isomer of 1BDD. Energy drop 21 kcal/mol. See text for details. First row shows the native sequence and the second row shows the energy-optimized sequence below:

QQNAFYEILHLPNLNEEQRNGFIQSLKDDPSQSANLLAEAKKLNDA
SPAEYKKEALDQAIQLSGDPESNFIKLQEALLLFNHNQQALRNNDN

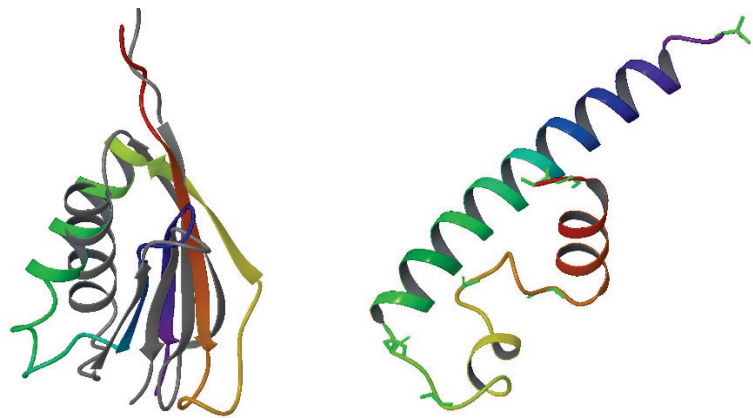

FIG. 4. Ground state sequence-permutation isomer of 1IGD. Energy drop 41 kcal/mol. See text for details. First row shows the native sequence and the second row shows the energy-optimized sequence below:

**GMTPAVTTYKLVINGKTLKGETTTKAVDAETAEKAFKQYANDNGVDGVWTYDDATKTFTVTEG**

**GGEAAVTMWANVGEVDGTYKFTDKADGKPATDTNTETANYVVQFILVAKKLTDTYETTKTKTG**

First, secondary structure prediction performed on the optimized sequences by multiple trusted web servers [20] provides no evidence for helical segments in either the 1IGD or the 1E0L ground state permutation isomers; the predictions are all beta and random coils (except for a tiny alpha contribution in 1IGD). Not surprisingly, however, 1BDD itself being an all-alpha protein, retains its alpha character after sequence optimization. Second, the extreme case of 1E0L where sequence optimization results in re-folding a pure triple-stranded antiparallel beta-sheet into a pure alpha helical structure, pleads for more accurate calculations. Here we employ multi-scale modeling by (i) generating the full backbone structure from the C-alpha coordinates [21], (ii) applying highly accurate side chain prediction using SCWRL4 [22], and finally (iii) running all-atom molecular dynamics (MD) simulation with explicit solvent. We carry out the MD simulation with Desmond [23] running on a NVIDIA GeForce GTX 780 graphics processor and using a simulation protocol that has proven highly successful in folding numerous small proteins [24]. The length of the MD simulation is 150 ns and the MPEG video file (70 MB) is available upon request. The trajectory provides vivid graphical evidence that the all-alpha UNRES ground-state sequence-permutation isomer of 1E0L is not stable at all, it entirely unfolds in less than 10 ns and remains in a highly volatile, flexible and somewhat U-shaped random coil configuration for the rest of the simulation.

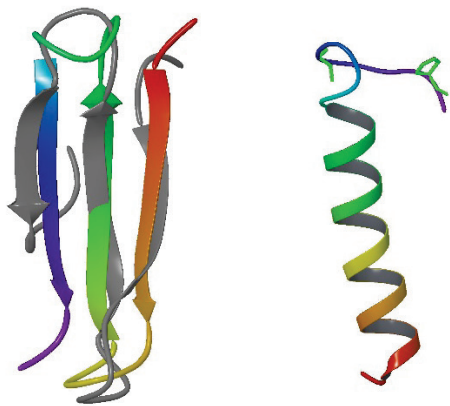

FIG. 5. Ground state sequence-permutation isomer of 1E0L. Energy drop 45 kcal/mol. See text for details. First row shows the native sequence and the second row shows the energy-optimized sequence below:

**SEWTEYKTADGKTYYYNNRTLESTWEKP**
**NRTETYLSNTETYEYEKKWTWADGSKPY**

IV. SUMMARY AND PERSPECTIVE

In this communication, we propose a dual concept for the global optimization of off-lattice protein models predicting optimal sequences as well as optimal folds and embed our approach in multi-scale modeling. Using hidden-force Monte Carlo optimization we find the optimal fold of simple and detailed protein models. Further, we also find the protein sequence that yields the lowest energy fold amongst all sequences for a given chain length and residue mixture. We then generate an all-atom model and run state-of-the-art MD simulation. In particular, we find that with a simplified potential, binary protein AB models fold into globular, compact structures with no tendency to form knots, and with a "hydrophobic" core. Historically, these models use a Fibonacci series sequence protein with chain lengths of 13, 21, 34, 55, and 89. We show that sequence optimization yields non-Fibonacci sequences that fold into still lower energy structures with more compact cores than the original Fibonacci sequences. We extend our investigation to study real proteins using the sophisticated UNRES model and find via sequence optimization and all-atom MD simulation that UNRES inherently favors helical structures.

We demonstrate that our methodology is applicable to detailed protein models, fits well within multi-scale modeling, and ultimately may aid de novo protein design by yielding novel folds. However, we also demonstrate that further improvements in force fields are needed before sequence optimization could become a reliable tool for proteins in general, and for the moment our research focus stays within the realm of all-alpha proteins, studying sequences that afford stable folds—whether native or non-native.

**Acknowledgement.** I am indebted to Adam Liwo for instrumental assistance with UNRES.

Title: Conceptual framework for performing simultaneous fold and sequence optimization in multi-scale protein modeling

Author: Istvan Kolossvary

Description:
List of coordinates, sequences, and energies of putative global minimum folds of PAB and UNRES models. The coordinates are Cartesian coordinates. The sequence is identified by the residue type A or B in the first column. Energies are absolute energies in epsilon units. The first 5 models were computed with the potential in Eq. 1. The other 5 models were computed with the simplified potential including the pseudo Lennard-Jones potential, the bond length restraint and the flat-bottom angle bending constraint to keep the Calpha-Calpha-Calpha bond angles within the 85-145 degree range. The UNRES protein models below are represented by their C-alpha atoms listed in PDB format. The energy is UNRES/4P.

S_13 Fibonacci sequence global minimum energy computed by Eq. 1 =
                                                                    -27.171382
```
A     -2.210729   -0.766557    0.859473
B     -1.468936   -0.515469    1.481321
B     -1.125457    0.413991    1.346684
A     -2.015611    0.307761    0.903583
B     -2.875024    0.042923    0.466239
B     -2.923165   -0.824726   -0.028602
A     -2.104003   -1.398285   -0.026841
B     -1.292453   -1.270099    0.543208
A     -1.100612   -0.289674    0.498904
B     -1.256479    0.656395    0.214902
B     -2.203998    0.733500   -0.095357
A     -1.998733   -0.244823   -0.122725
B     -1.256899   -0.823838   -0.460991
```

S_21 Fibonacci sequence global minimum energy computed by Eq. 1 =
                                                                    -56.408990
```
B     -3.502088   -2.347106   -2.580986
A     -3.633583   -3.070794   -1.903502
B     -3.182330   -2.964804   -1.017422
A     -2.558379   -2.207111   -0.826145
B     -1.942815   -1.419028   -0.828706
B     -1.631517   -1.022569   -1.692370
A     -1.917468   -1.469613   -2.539947
B     -2.396659   -2.287089   -2.859494
A     -2.809866   -3.188090   -2.727356
B     -2.658749   -3.616882   -1.836679
B     -2.088394   -3.137323   -1.169805
A     -1.497975   -2.495615   -0.680298
B     -0.926488   -1.760655   -1.045314
B     -0.814412   -1.662030   -2.034107
A     -1.235498   -2.340732   -2.635811
B     -1.870827   -3.002000   -2.236963
A     -2.685908   -2.555420   -1.867897
B     -3.446519   -2.008019   -1.518868
B     -2.589028   -1.522775   -1.689879
```

```
A    -1.763180   -2.085958   -1.661572
B    -0.995117   -2.724986   -1.620079
```

S_34 Fibonacci sequence global minimum energy computed by Eq. 1 =
                                                       -106.115214
```
A    -0.054962    2.229487   -0.900343
B    -0.853011    2.143012   -1.496697
B    -0.951238    1.732429   -0.590181
A    -0.524775    1.446744    0.268022
B     0.330605    1.385370    0.782374
B     1.265892    1.429216    1.133538
A     1.199972    1.905059    0.256483
B     0.995707    2.354906   -0.612946
A     0.707322    2.800658   -1.460375
B    -0.237086    3.129292   -1.450697
B    -0.878686    2.917563   -0.713459
A    -0.848802    2.520503    0.203846
B    -0.411806    2.204694    1.046045
B     0.565968    2.364524    1.181737
A     1.553768    2.506594    1.117971
B     2.284360    2.032034    0.627028
A     2.006677    1.145438    0.257113
B     1.079417    0.771783    0.233288
B     0.127003    0.470460    0.187356
A    -0.495836    0.717894   -0.554834
B    -0.200073    1.245807   -1.350967
A     0.708342    1.655607   -1.433690
B     1.614781    2.075060   -1.482927
B     1.824123    2.954070   -1.054539
A     1.063933    3.448186   -0.632687
B     0.147975    3.147931   -0.366482
B     0.176094    2.278401    0.126597
A     0.388799    1.417582   -0.335729
B     0.682145    0.618698   -0.860826
A     1.419258    1.273341   -0.693175
B     2.059386    2.006824   -0.464622
B     1.809237    2.833024    0.040171
A     0.893430    3.031008    0.389595
B    -0.039220    3.183399    0.716612
```

S_55 Fibonacci sequence global minimum energy computed by Eq. 1 =
                                                       -190.579174
```
B    -1.048644   -1.434770   -2.612892
A    -0.857230   -1.125885   -1.681255
B    -0.596707   -0.675295   -0.827388
A    -0.290393   -0.155407   -0.029967
B     0.110010    0.484079    0.626335
B     0.162770    0.776914   -0.328370
A     0.162639    0.678281   -1.323493
B     0.219978    0.432508   -2.291121
A     0.444364    0.055716   -3.189829
B     1.268189   -0.503306   -3.283678
B     1.459310   -1.259428   -2.657776
A     0.965495   -1.656927   -1.884379
```

```
B         0.255341    -1.879653    -1.216491
B        -0.653161    -1.783533    -0.809818
A        -1.504582    -1.262869    -0.746650
B        -1.578376    -0.369291    -1.189448
A        -1.557596     0.523631    -1.639172
B        -0.739898     1.066675    -1.830141
B         0.113194     1.559915    -2.000266
A         1.027615     1.166839    -1.903716
B         1.928092     0.747291    -1.789189
A         2.048504     0.042246    -1.090328
B         1.410687    -0.255968    -0.380218
B         0.694872    -0.476615     0.282294
A        -0.074711    -0.632197     0.901598
B        -0.858279    -0.013484     0.844884
B        -0.934677     0.677467     0.126029
A        -0.778999     0.418222    -0.827153
B        -0.560743    -0.043843    -1.686720
A        -0.310408    -0.533405    -2.521979
B         0.337372    -1.040818    -3.090234
B         0.770256    -0.531522    -2.346440
A         1.031028     0.081018    -1.600259
B         1.181685     0.761207    -0.882872
B         0.607766     1.578979    -0.926071
A        -0.391078     1.531446    -0.918889
B        -1.367073     1.318376    -0.873801
A        -1.829006     0.501407    -0.528579
B        -1.397912    -0.309658    -0.133193
B        -0.837781    -1.075151     0.183464
A         0.074510    -1.212677    -0.202294
B         0.924085    -1.176072    -0.728487
A         1.540988    -0.860680    -1.449567
B         1.854634    -0.317133    -2.228143
B         1.320831     0.424115    -2.635087
A         0.699667     1.130165    -2.975156
B        -0.300051     1.108245    -2.966151
B        -0.904839     0.382850    -2.637471
A        -1.426102    -0.387277    -2.269794
B        -1.917393    -1.171891    -1.891624
A        -1.532661    -2.027320    -1.544895
B        -0.607225    -2.241704    -1.857312
B        -0.015642    -1.517680    -2.212015
A         0.222705    -0.833229    -1.523018
B         0.419780    -0.197002    -0.777114
```

S_89 Fibonacci sequence global minimum energy computed by Eq. 1 = -325.578412

```
A        -0.890676     5.222297    11.293380
B        -1.668704     4.931778    10.736363
B        -2.158498     5.528157    10.100414
A        -1.668164     6.004270     9.370424
B        -1.051123     6.389643     8.684315
B        -0.152088     6.170183     8.305408
A         0.750047     5.925282     7.950197
B         1.538695     5.358181     8.187744
```

```
A     0.826827    4.723130    7.887825
B     0.036563    4.606234    7.286312
B    -0.894691    4.884079    7.050588
A    -1.682813    4.670039    7.627694
B    -2.347825    4.340134    8.297708
B    -2.648735    3.820308    9.097229
A    -2.285703    3.421407    9.939301
B    -1.657824    3.205639   10.687106
A    -1.047698    2.438666   10.488372
B    -0.448592    1.679773   10.233127
B    -0.073118    1.635437    9.307356
A     0.175184    2.320731    8.622727
B     0.429502    3.028740    7.963908
A     0.961370    3.773140    8.367617
B     1.461699    4.496864    8.842896
B     1.060899    4.700943    9.736040
A     0.580722    4.800837   10.607502
B     0.051615    4.760676   11.455104
B    -0.206286    3.937065   11.960229
A     0.307527    3.094937   11.796475
B     0.935780    2.921734   11.037992
A     1.503541    2.870146   10.216418
B     2.004594    2.923607    9.352655
B     1.989546    3.553174    8.575855
A     1.817329    4.208172    7.840111
B     1.556098    4.869972    7.137416
B     0.721844    5.407230    7.013423
A    -0.183812    5.670363    7.345912
B    -0.999588    5.548707    7.911342
A    -1.748901    5.296576    8.523683
B    -2.477520    5.020295    9.150407
B    -2.337992    4.467211    9.971765
A    -1.423481    4.092965   10.125423
B    -0.656610    4.211696    9.494699
A     0.233414    4.007857    9.902503
B     1.076682    3.845623   10.414925
B     1.784226    4.498139   10.686219
A     1.493482    5.421480   10.435407
B     0.703605    5.818557    9.968050
B    -0.134302    6.070308    9.483763
A    -0.743820    5.424754    9.023605
B    -0.914901    4.602955    8.480119
A    -0.076806    5.105053    8.266837
B     0.622976    5.476910    8.876777
B     1.522606    5.563877    9.304680
A     2.156608    4.894567    9.692060
B     1.890921    3.930661    9.709134
A     1.058479    3.482964    9.382628
B     0.138880    3.359732    9.009600
B    -0.812989    3.507476    8.741054
A    -1.610013    4.038142    9.029405
B    -1.492645    4.909885    9.505099
B    -0.979113    5.537814   10.089894
A    -0.136436    5.670471   10.611711
```

```
B        0.713138     5.769108    11.129872
A        1.235614     4.994903    11.487115
B        0.879173     4.060589    11.489658
B        0.131152     3.711345    10.925309
A       -0.587034     3.420047    10.293368
B       -1.290857     3.148875     9.636790
B       -1.913205     2.998510     8.868630
A       -1.660815     3.509169     8.046727
B       -0.894610     3.973606     7.602627
A       -0.054379     4.003275     8.144040
B        0.339086     4.438247     8.953967
B        0.011977     5.023939     9.695563
A       -0.501837     4.623321    10.454180
B       -0.959711     4.067838    11.148288
A       -0.714692     3.117182    11.338593
B       -0.066216     2.493851    10.901628
B        0.583538     2.018076    10.308790
A        0.927769     2.356424     9.432990
B        1.249645     2.721486     8.559421
B        1.494238     3.149799     7.689523
A        0.878466     3.814968     7.267178
B       -0.065908     3.506738     7.152503
A       -0.606076     2.976902     7.806334
B       -0.954466     2.474637     8.597758
B       -1.161863     2.043503     9.475885
A       -0.346425     2.607860     9.604576
B        0.415302     3.024958    10.100354
```

S_13 Fibonacci sequence global minimum energy computed by LJ term =
                                                       -18.264451

```
A       -1.872515    -0.893300     1.154829
B       -2.731132    -0.568108     0.758563
B       -2.520868     0.363275     1.055738
A       -1.590969     0.182426     1.376019
B       -1.309534     0.775524     0.621679
B       -0.785642     0.122565     0.074705
A       -0.850686    -0.469205     0.878182
B       -1.174009    -1.161829     0.233409
A       -2.165186    -1.224768     0.116772
B       -2.796179    -0.659337    -0.414391
B       -2.608159     0.249817    -0.042793
A       -1.790042    -0.174069     0.345796
B       -1.637280    -0.521693    -0.579308
```

S_21 Fibonacci sequence global minimum energy computed by LJ term =
                                                       -38.280724

```
B       -2.598970    -1.071744    -0.989501
A       -2.555464    -1.832540    -1.637030
B       -3.179306    -2.381891    -1.081124
A       -3.006899    -3.100627    -1.754692
B       -2.258990    -3.762954    -1.710463
B       -1.448311    -3.577823    -1.155012
A       -1.257632    -2.619644    -0.941625
B       -0.956978    -2.301922    -1.840880
```

```
A      -1.693473   -1.901240   -2.385884
B      -1.498008   -2.638748   -3.032317
B      -2.233398   -3.230214   -2.701607
A      -2.734630   -2.377477   -2.554629
B      -3.551075   -2.074591   -2.063023
B      -3.307601   -1.117647   -1.904976
A      -2.600028   -1.274972   -2.593879
B      -1.969334   -0.931424   -1.898036
A      -1.564488   -1.608399   -1.283378
B      -2.179463   -2.041624   -0.624500
B      -2.319442   -3.001169   -0.868793
A      -1.967465   -2.733347   -1.765666
B      -1.265945   -3.308704   -2.186183

S_34 Fibonacci sequence global minimum energy computed by LJ term =
                                                       -75.170999
A       0.472359    2.418285   -0.638095
B      -0.322102    2.299673   -1.233713
B      -0.047300    1.650925   -1.943368
A       0.691744    1.410112   -1.314225
B       0.478766    0.496016   -0.969161
B      -0.167306    0.439314   -0.207994
A      -0.553580    1.238591    0.252382
B       0.290259    1.228340    0.788878
A       0.371417    2.187270    0.517094
B       0.329232    2.937186    1.177280
B       1.087189    3.424014    0.743116
A       1.236341    2.671553    0.101590
B       2.127574    2.239059    0.238152
B       2.271682    2.431017   -0.732612
A       1.412721    2.888905   -0.961792
B       0.950592    3.588052   -0.416237
A       0.206155    3.177982    0.110693
B      -0.407799    3.236765   -0.676456
B       0.371919    3.267500   -1.301832
A       0.730286    2.427743   -1.709729
B       1.608837    1.981712   -1.538835
A       1.389475    1.818740   -0.576900
B       1.491686    0.851222   -0.808113
B       0.968414    0.654887    0.021127
A       0.390090    1.417616   -0.268326
B      -0.318322    1.226465   -0.947748
B      -1.130292    1.717089   -0.631537
A      -0.481968    2.287088   -0.126787
B      -0.645757    2.798181    0.716986
A      -0.449328    1.931999    1.176489
B       0.464218    1.940154    1.583146
B       1.267028    2.344961    1.145395
A       1.267005    1.595162    0.483729
B       2.057765    1.152922    0.060503
```

S_55 Fibonacci sequence global minimum energy computed by LJ term =
                                                          -137.241642

| | | | |
|---|---:|---:|---:|
| B |  1.373813 |  0.169900 | -0.003080 |
| A |  0.839574 |  0.156997 | -0.848314 |
| B |  1.224370 |  0.867832 | -1.437077 |
| A |  0.377357 |  0.770842 | -1.959722 |
| B | -0.189950 |  1.593290 | -2.001428 |
| B | -0.752958 |  1.436535 | -1.189978 |
| A | -0.408800 |  1.311824 | -0.259385 |
| B | -1.308185 |  0.916183 | -0.073446 |
| A | -0.858252 |  0.037826 | -0.234832 |
| B | -1.284447 | -0.789417 |  0.131262 |
| B | -0.695297 | -1.273802 | -0.515477 |
| A | -0.006945 | -0.549511 | -0.555161 |
| B |  0.858993 | -0.783863 | -0.113315 |
| B |  0.673085 | -0.297359 |  0.740354 |
| A | -0.288062 | -0.478856 |  0.532380 |
| B |  0.025050 | -1.402222 |  0.310224 |
| A |  0.334731 | -1.623659 | -0.614471 |
| B | -0.271404 | -2.223190 | -1.137120 |
| B | -1.125174 | -1.780380 | -1.410972 |
| A | -1.487109 | -1.038236 | -1.975084 |
| B | -1.530026 | -0.039385 | -1.996363 |
| A | -0.665714 |  0.460180 | -1.938074 |
| B | -1.205005 |  1.252159 | -2.224309 |
| B | -1.998532 |  0.927871 | -1.709381 |
| A | -1.323163 |  0.539623 | -1.082374 |
| B | -1.904416 | -0.024012 | -0.495467 |
| B | -1.661492 | -0.902707 | -0.906419 |
| A | -0.844182 | -0.435675 | -1.243885 |
| B | -0.228644 | -1.185260 | -1.487269 |
| A |  0.415769 | -0.978629 | -2.223498 |
| B |  0.308049 | -1.404688 | -3.121756 |
| B | -0.235283 | -0.595015 | -3.343612 |
| A |  0.259806 |  0.006437 | -2.716602 |
| B |  1.061168 | -0.536598 | -2.967464 |
| B |  1.955183 | -0.345303 | -2.562321 |
| A |  1.681263 | -1.004307 | -1.861835 |
| B |  1.202204 | -1.648374 | -2.458224 |
| A |  0.675894 | -1.879328 | -1.639899 |
| B | -0.125260 | -1.963959 | -2.232342 |
| B | -0.771985 | -1.419525 | -2.766510 |
| A | -0.580275 | -0.552099 | -2.307364 |
| B | -0.939675 |  0.136461 | -2.937217 |
| A | -0.345111 |  0.937011 | -2.862293 |
| B |  0.600021 |  1.261688 | -2.898470 |
| B |  1.277340 |  0.561780 | -2.671830 |
| A |  1.081183 | -0.097319 | -1.945809 |
| B |  1.974153 |  0.069299 | -1.527669 |
| B |  1.666049 | -0.570189 | -0.823307 |
| A |  0.795558 | -0.858027 | -1.222548 |
| B |  0.130805 | -0.203018 | -1.581805 |
| A | -0.201357 |  0.515911 | -0.971229 |
| B |  0.326022 |  1.360430 | -1.064280 |
| B | 0.837993 | 1.095576 | -0.247129 |
| A | 0.201303 | 0.373024 | 0.022210 |
| B | -0.534534 | 0.612033 | 0.655785 |

S_89 Fibonacci sequence global minimum energy computed by LJ term = 
                                                        -242.825509

| | | | |
|---|---:|---:|---:|
| A | 0.262401 | 4.242779 | 8.095651 |
| B | 1.068432 | 4.121103 | 7.516420 |
| B | 0.659709 | 3.332733 | 7.056616 |
| A | -0.176719 | 3.349537 | 7.604434 |
| B | -1.092642 | 3.300990 | 7.206027 |
| B | -0.785156 | 4.044326 | 6.611969 |
| A | 0.099885 | 4.260447 | 7.024268 |
| B | 0.580675 | 5.096724 | 7.287862 |
| A | 1.005547 | 5.043997 | 8.191578 |
| B | 1.610226 | 4.300315 | 8.476709 |
| B | 1.707206 | 3.305931 | 8.519077 |
| A | 0.790503 | 3.287169 | 8.119952 |
| B | 0.917618 | 2.416369 | 7.645033 |
| B | -0.020586 | 2.336864 | 7.308208 |
| A | -0.820510 | 2.488857 | 7.888742 |
| B | -1.047035 | 3.375591 | 8.291715 |
| A | -0.135793 | 3.408852 | 8.702235 |
| B | 0.110602 | 2.490435 | 8.392721 |
| B | 0.127264 | 1.999318 | 9.263654 |
| A | -0.024691 | 2.898665 | 9.673627 |
| B | -0.906480 | 3.115770 | 10.092323 |
| A | -1.040584 | 3.705501 | 9.295941 |
| B | -1.951092 | 3.357325 | 9.072896 |
| B | -1.708706 | 2.545623 | 8.541497 |
| A | -0.829295 | 2.606496 | 9.013652 |
| B | -0.838998 | 2.075203 | 9.860784 |
| B | -0.099781 | 2.182613 | 10.525629 |
| A | -0.039922 | 3.167379 | 10.688878 |
| B | -0.649078 | 3.759855 | 11.216037 |
| A | -0.886562 | 4.157247 | 10.329651 |
| B | -1.805176 | 3.938417 | 10.000617 |
| B | -2.328988 | 4.370486 | 9.266494 |
| A | -1.628394 | 4.257806 | 8.561887 |
| B | -1.998209 | 3.625284 | 7.881336 |
| B | -1.661353 | 4.286907 | 7.211425 |
| A | -0.799669 | 4.226626 | 7.715275 |
| B | -0.551235 | 5.086298 | 7.268911 |
| A | -0.073144 | 5.290318 | 8.123196 |
| B | 0.571077 | 6.037373 | 7.959217 |
| B | 0.517272 | 6.824111 | 8.574157 |
| A | 0.401977 | 6.758480 | 9.565318 |
| B | 1.204549 | 6.162898 | 9.531305 |
| A | 0.617416 | 5.492169 | 9.984507 |
| B | 0.713704 | 6.080373 | 10.787467 |
| B | -0.234202 | 6.389185 | 10.865619 |
| A | -0.345923 | 5.994866 | 9.953464 |
| B | -0.724722 | 6.652212 | 9.302000 |
| B | -0.402554 | 6.266276 | 8.437558 |

```
A       0.307962    5.836758    8.994945
B       0.333013    4.837120    9.004377
A      -0.571119    4.495350    8.747993
B      -1.120092    5.123399    8.196463
B      -1.849419    5.310010    8.854684
A      -1.339882    4.756065    9.513103
B      -2.071029    4.998725   10.150706
A      -1.718019    4.498096   10.941120
B      -0.872181    4.798858   11.381687
B       0.077675    4.558385   11.581548
A       0.182024    4.230095   10.642753
B       1.108881    3.982703   10.925125
B       1.911000    4.295324   10.416331
A       1.632244    5.247700   10.292738
B       0.902394    5.026502   10.939569
A      -0.052887    5.286314   10.798382
B      -0.303312    5.670244   11.687135
B      -1.181011    5.874377   11.253575
A      -1.080298    5.286839   10.450671
B      -1.417309    5.856847    9.701329
B      -0.703314    5.492788    9.103273
A      -0.296294    4.909130    9.805896
B      -0.038124    3.964528    9.603241
A       0.775288    3.919465    9.023301
B       1.664446    3.811010    9.467861
B       2.195119    4.657098    9.417666
A       1.314974    5.095290    9.235099
B       0.926354    4.483003    9.923632
A       0.806105    3.497458   10.042964
B       1.700779    3.211873   10.386473
B       1.777215    2.716025    9.521436
A       0.879318    2.857280    9.104515
B       1.042755    1.975180    8.662719
B       1.157882    1.628993    9.593793
A       0.871987    2.423081   10.130158
B       0.909424    2.861616   11.028093
A       0.419355    3.524880   11.593702
B      -0.266002    2.803327   11.691918
B      -0.983369    2.604411   11.024224
A      -1.597429    3.379182   10.873698
B      -1.932869    2.849475   10.094665
```

S_13 Optimized sequence global minimum energy computed by LJ term =
                                                        -18.610603
```
A       1.038830   -0.765684    0.479799
A       1.015785   -0.076582   -0.244499
B       0.669287   -0.248282   -1.166702
B      -0.056089   -0.909581   -0.975615
A       0.055459   -0.625012   -0.023471
B       0.255497   -1.586615    0.164427
B       1.221177   -1.841911    0.212231
B       1.814616   -1.696593   -0.579420
A       1.903484   -0.763329   -0.231396
B       1.786908   -0.246508   -1.079514
```

| | | | |
|---|---|---|---|
| B | 1.376632 | -1.087295 | -1.432712 |
| A | 0.902306 | -1.121712 | -0.553038 |
| B | 0.741207 | -2.082997 | -0.776590 |

S_21 Optimized sequence global minimum energy computed by LJ term =
-39.650163

| | | | |
|---|---|---|---|
| B | -1.168606 | -0.122705 | -2.808934 |
| A | -1.862785 | 0.435824 | -2.354887 |
| A | -2.028844 | 0.286177 | -1.380195 |
| B | -2.208717 | 0.787195 | -0.533660 |
| B | -2.321778 | 1.581406 | -1.130690 |
| A | -2.688791 | 1.008135 | -1.863261 |
| B | -3.071911 | 0.417466 | -1.153099 |
| A | -2.710029 | -0.513141 | -1.098203 |
| B | -1.740871 | -0.628266 | -0.880305 |
| B | -1.228007 | 0.201424 | -0.659886 |
| B | -1.321287 | 1.055575 | -1.171476 |
| B | -1.642009 | 1.469229 | -2.023548 |
| B | -2.236188 | 1.350234 | -2.819030 |
| A | -2.846012 | 0.559679 | -2.875036 |
| A | -2.806989 | -0.083118 | -2.109995 |
| A | -1.972821 | -0.621171 | -1.988909 |
| B | -1.118503 | -0.184810 | -1.706544 |
| B | -0.832964 | 0.693723 | -2.089489 |
| B | -1.179475 | 1.016090 | -2.970403 |
| B | -1.880494 | 0.483777 | -3.444972 |
| A | -2.235819 | -0.357323 | -3.037179 |

S_34 Optimized sequence global minimum energy computed by LJ term =
-77.919140

| | | | |
|---|---|---|---|
| A | -1.545522 | 1.177180 | 0.136790 |
| B | -1.852461 | 0.277788 | -0.174471 |
| B | -2.305021 | -0.428425 | 0.370003 |
| B | -3.101351 | 0.175172 | 0.330931 |
| A | -3.195326 | 0.957649 | 0.946476 |
| B | -4.038946 | 1.291080 | 1.367341 |
| B | -4.224617 | 1.700988 | 0.474313 |
| A | -3.255779 | 1.948018 | 0.456213 |
| B | -2.867313 | 2.642656 | -0.149237 |
| A | -2.003360 | 2.160413 | -0.004241 |
| A | -1.353174 | 2.093840 | 0.752612 |
| B | -0.702807 | 1.334230 | 0.756818 |
| B | -1.148185 | 0.439531 | 0.790805 |
| A | -2.144132 | 0.521769 | 0.827252 |
| B | -2.763715 | 0.018408 | 1.429532 |
| B | -3.745329 | 0.034185 | 1.239312 |
| B | -4.162410 | 0.586755 | 0.517710 |
| B | -3.599110 | 1.065813 | -0.155487 |
| A | -2.615162 | 1.147361 | 0.003235 |
| A | -2.300719 | 1.565377 | 0.855517 |
| A | -1.531879 | 1.230223 | 1.400088 |
| B | -1.698313 | 0.324516 | 1.789952 |
| B | -2.423976 | 0.230089 | 2.471492 |
| B | -2.808893 | 1.058299 | 2.878807 |

```
B   -2.591776    1.960076    2.505099
A   -2.020424    2.161472    1.709489
A   -2.436276    2.622496    0.925573
B   -3.342021    3.018557    0.774705
B   -3.946384    2.397500    1.273737
A   -3.103517    1.925973    1.533034
B   -3.622960    1.621311    2.331381
B   -3.408948    0.689301    2.038883
A   -2.513309    1.063787    1.798910
B   -1.783385    1.210412    2.466526
```

S_55 Optimized sequence global minimum energy computed by LJ term = -146.316053

```
A   -2.973835    0.271336   -0.744174
A   -2.379573   -0.505505   -0.535926
B   -2.754117   -1.430975   -0.592679
B   -3.648686   -1.649556   -0.202848
B   -3.695617   -1.228583    0.703010
A   -3.529803   -0.268431    0.927986
B   -3.858544    0.200964    1.747496
B   -4.795079    0.534182    1.856442
B   -5.226262    1.098880    1.152737
B   -5.657353    0.725177    0.331451
B   -5.332960   -0.127324   -0.078441
A   -4.412782   -0.344033    0.247606
B   -4.467569   -0.661575    1.194267
B   -3.727868   -0.898194    1.824230
B   -2.825390   -0.967957    1.399183
B   -2.667484   -0.976216    0.411764
A   -3.408784   -0.604395   -0.147004
A   -3.362471   -0.704323   -1.140919
B   -2.573241   -0.313214   -1.614362
B   -1.827804    0.182300   -1.168505
A   -2.052405    0.686429   -0.334595
B   -1.634234   -0.087315    0.141275
B   -1.988619   -0.352474    1.037991
B   -2.719519    0.142977    1.507366
A   -3.266001    0.829831    1.028211
A   -2.887597    1.186242    0.173939
B   -3.195715    2.069915   -0.178455
B   -3.631644    2.059588   -1.078376
B   -3.574188    1.270052   -1.689385
A   -3.540737    0.270685   -1.677332
A   -4.301702   -0.377270   -1.710315
A   -5.072404   -0.159082   -1.111640
B   -5.432354    0.717969   -0.793489
A   -4.650887    0.712986   -0.169563
B   -5.035067    1.568016    0.178760
B   -4.598191    2.046878    0.940226
B   -4.142720    1.313925    1.445523
A   -4.369110    0.524691    0.874686
B   -5.225473    0.021948    0.992555
B   -5.377799   -0.910216    0.664122
B   -4.656841   -1.377197    0.152117
```

| | | | |
|---|---|---|---|
| A | -4.358333 | -0.925402 | -0.688582 |
| A | -4.064674 | 0.028497 | -0.750527 |
| A | -4.475313 | 0.734922 | -1.327012 |
| B | -4.626157 | 1.593895 | -0.837716 |
| B | -4.254866 | 2.224542 | -0.156227 |
| A | -3.969040 | 1.409887 | 0.348391 |
| B | -3.388986 | 1.976633 | 0.933481 |
| B | -2.482623 | 1.596761 | 1.118423 |
| B | -2.082081 | 0.738179 | 0.798419 |
| A | -2.703822 | 0.106200 | 0.335778 |
| A | -3.627577 | 0.416598 | 0.111441 |
| A | -3.686799 | 1.075113 | -0.638791 |
| B | -2.761678 | 1.354487 | -0.895892 |
| B | -2.574822 | 0.783251 | -1.695125 |

S_89 Optimized sequence global minimum energy computed by LJ term = -261.401077

| | | | |
|---|---|---|---|
| A | 5.335687 | 2.905267 | -9.716160 |
| A | 5.832940 | 3.467213 | -9.055139 |
| A | 6.573647 | 2.963723 | -8.610340 |
| A | 6.588762 | 1.996723 | -8.356021 |
| B | 5.847304 | 1.652888 | -7.779812 |
| B | 5.450839 | 0.761876 | -8.000976 |
| B | 5.614742 | 0.326729 | -8.886291 |
| B | 6.455216 | 0.556758 | -9.376892 |
| A | 6.901018 | 1.437602 | -9.217604 |
| A | 7.452963 | 1.296742 | -8.395709 |
| A | 8.130765 | 1.874034 | -8.851024 |
| A | 7.343220 | 2.415189 | -9.145851 |
| A | 7.573687 | 3.296985 | -8.734368 |
| B | 8.399980 | 3.828942 | -8.919462 |
| B | 7.784366 | 4.611600 | -9.011471 |
| B | 6.863774 | 4.955809 | -8.826990 |
| B | 5.981103 | 4.509978 | -8.678237 |
| B | 6.107616 | 3.756833 | -8.032664 |
| B | 5.201538 | 3.341739 | -8.114643 |
| B | 4.914555 | 2.436884 | -7.800192 |
| B | 4.961990 | 1.641023 | -8.403811 |
| A | 5.822732 | 1.383511 | -8.842915 |
| B | 6.484006 | 0.839015 | -8.326934 |
| B | 6.836661 | 1.268568 | -7.495599 |
| B | 6.784960 | 2.248266 | -7.301905 |
| B | 6.008944 | 2.723401 | -7.716685 |
| A | 5.606019 | 2.472408 | -8.596829 |
| A | 6.184722 | 2.264265 | -9.385352 |
| A | 6.065528 | 1.427481 | -9.919749 |
| B | 5.418240 | 0.665944 | -9.952566 |
| B | 4.790851 | 0.986499 | -9.242901 |
| A | 5.132237 | 1.897241 | -9.475287 |
| B | 4.599280 | 2.534158 | -8.918247 |
| B | 4.710900 | 3.517102 | -9.064396 |
| B | 5.232536 | 4.180442 | -9.600936 |
| A | 5.942526 | 3.752122 | -10.159911 |
| A | 6.116817 | 3.564621 | -11.126589 |

```
B     5.343402     3.660016    -11.753270
B     5.037800     2.712679    -11.848961
A     5.925344     2.478301    -11.452313
B     6.425150     3.088922    -12.066588
B     6.376102     4.087309    -12.037999
B     5.764629     4.603617    -11.438394
B     6.511205     4.548973    -10.775344
A     6.987514     3.704822    -10.529306
B     7.819385     4.211375    -10.756005
B     8.393817     4.441852     -9.970570
B     8.697227     3.490507    -10.024274
A     8.324763     2.816763     -9.386044
B     8.431279     2.794482     -8.391982
A     7.559624     2.419576     -8.076287
B     7.090471     3.228031     -7.720905
B     7.129588     4.108302     -8.193762
A     6.794454     3.948602     -9.122298
A     6.713603     3.124316     -9.682663
A     6.139630     2.723901    -10.396961
A     5.267798     3.014540    -10.791219
B     4.503855     2.386220    -10.938199
B     4.323784     2.515777     -9.963116
B     4.521673     3.479175    -10.143959
B     5.075225     4.062989    -10.737880
B     5.536667     4.879314    -10.390486
B     6.245523     4.684382     -9.712604
B     7.218278     4.576227     -9.917663
A     7.676200     3.711366     -9.711939
A     7.613535     2.857795    -10.229128
A     6.959172     2.114544    -10.089911
A     6.478405     1.761216    -10.892414
B     6.988109     1.005053    -11.302800
B     7.362086     0.196718    -10.848117
B     7.433998     0.151774     -9.851718
A     7.851979     0.980870     -9.480380
A     7.972920     1.876910     -9.907562
B     8.564523     2.432906    -10.491407
B     8.228152     3.223417    -11.003215
B     7.833432     2.633804    -11.707879
A     7.642734     1.991391    -10.965627
B     7.933594     1.092223    -10.638680
A     7.025067     1.046270    -10.223394
B     6.376887     0.309481    -10.415756
B     5.815884     0.956523    -10.932099
B     4.839685     0.739740    -10.938399
B     4.591087     1.453429    -10.283537
A     5.409067     1.958511    -10.558853
B     5.186689     1.664048    -11.488282
B     6.093856     1.482577    -11.867911
B     6.892296     2.082913    -11.822188
A     6.928889     2.779607    -11.105753
B     7.273290     3.603558    -11.555743
```

```
1BDD
REMARK    E=     -600.67000  Rg=    2.942  SEQ=
QQNAFYEILHLPNLNEEQRNGFIQSLKDDPSQSANLLAEAKKLNDA
ATOM        1  CA  GLN     1       3.800   0.000   0.000       0.000
ATOM        2  CA  GLN     2       3.910  -3.798   0.000       0.000
ATOM        3  CA  ASN     3       2.071  -3.958  -3.321       0.000
ATOM        4  CA  ALA     4       5.026  -2.468  -5.189       0.000
ATOM        5  CA  PHE     5       7.282  -5.270  -3.966       0.000
ATOM        6  CA  TYR     6       4.770  -7.927  -5.000       0.000
ATOM        7  CA  GLU     7       4.137  -6.224  -8.337       0.000
ATOM        8  CA  ILE     8       7.818  -5.337  -8.659       0.000
ATOM        9  CA  LEU     9       8.917  -8.854  -7.732       0.000
ATOM       10  CA  HIS    10       6.631 -10.416 -10.335       0.000
ATOM       11  CA  LEU    11       7.421 -14.129 -10.173       0.000
ATOM       12  CA  PRO    12       4.717 -16.796 -10.045       0.000
ATOM       13  CA  ASN    13       5.064 -17.126  -6.275       0.000
ATOM       14  CA  LEU    14       4.320 -13.433  -5.784       0.000
ATOM       15  CA  ASN    15       1.445 -13.591  -8.265       0.000
ATOM       16  CA  GLU    16       0.065 -16.676  -6.528       0.000
ATOM       17  CA  GLU    17       0.778 -15.102  -3.144       0.000
ATOM       18  CA  GLN    18      -0.955 -11.904  -4.243       0.000
ATOM       19  CA  ARG    19      -4.027 -13.870  -5.309       0.000
ATOM       20  CA  ASN    20      -4.389 -15.441  -1.868       0.000
ATOM       21  CA  GLY    21      -3.735 -12.542   0.501       0.000
ATOM       22  CA  PHE    22      -0.100 -13.473   1.102       0.000
ATOM       23  CA  ILE    23       0.798  -9.982   2.306       0.000
ATOM       24  CA  GLN    24       0.083 -10.849   5.936       0.000
ATOM       25  CA  SER    25       2.002 -14.114   5.625       0.000
ATOM       26  CA  LEU    26       4.650 -12.508   3.425       0.000
ATOM       27  CA  LYS    27       4.868  -9.509   5.748       0.000
ATOM       28  CA  ASP    28       4.984 -11.763   8.805       0.000
ATOM       29  CA  ASP    29       7.472 -14.151   7.210       0.000
ATOM       30  CA  PRO    30       9.863 -11.440   6.038       0.000
ATOM       31  CA  SER    31       8.963  -9.279   3.044       0.000
ATOM       32  CA  GLN    32      12.602  -8.384   2.416       0.000
ATOM       33  CA  SER    33      13.562 -12.061   2.426       0.000
ATOM       34  CA  ALA    34      10.635 -12.926   0.162       0.000
ATOM       35  CA  ASN    35      11.369  -9.894  -2.007       0.000
ATOM       36  CA  LEU    36      15.069 -10.760  -2.004       0.000
ATOM       37  CA  LEU    37      14.261 -14.466  -2.235       0.000
ATOM       38  CA  ALA    38      11.600 -13.778  -4.860       0.000
ATOM       39  CA  GLU    39      13.869 -11.235  -6.541       0.000
ATOM       40  CA  ALA    40      16.883 -13.455  -5.890       0.000
ATOM       41  CA  LYS    41      14.888 -16.512  -6.948       0.000
ATOM       42  CA  LYS    42      13.582 -14.666 -10.002       0.000
ATOM       43  CA  LEU    43      16.977 -13.062 -10.588       0.000
ATOM       44  CA  ASN    44      18.733 -16.411 -10.210       0.000
ATOM       45  CA  ASP    45      16.533 -18.094 -12.811       0.000
ATOM       46  CA  ALA    46      18.211 -16.279 -15.698       0.000
```

```
1GAB
REMARK    E=    -680.97900  Rg=   2.382  SEQ=
LLKNAKEDAIAELKKAGITSDFYFNAINKAKTVEEVNALKNEILKAH
ATOM       1   CA   LEU     1      3.800    0.000    0.000    0.000
ATOM       2   CA   LEU     2      3.945   -3.797    0.000    0.000
ATOM       3   CA   LYS     3      0.795   -4.077   -2.106    0.000
ATOM       4   CA   ASN     4      2.780   -4.504   -5.318    0.000
ATOM       5   CA   ALA     5      5.485   -6.574   -3.634    0.000
ATOM       6   CA   LYS     6      2.894   -8.505   -1.635    0.000
ATOM       7   CA   GLU     7      0.757   -9.029   -4.733    0.000
ATOM       8   CA   ASP     8      3.821  -10.002   -6.761    0.000
ATOM       9   CA   ALA     9      5.372  -11.761   -3.771    0.000
ATOM      10   CA   ILE    10      2.027  -13.341   -2.901    0.000
ATOM      11   CA   ALA    11      1.446  -14.208   -6.555    0.000
ATOM      12   CA   GLU    12      5.006  -15.514   -6.807    0.000
ATOM      13   CA   LEU    13      4.671  -17.056   -3.350    0.000
ATOM      14   CA   LYS    14      1.267  -18.459   -4.292    0.000
ATOM      15   CA   LYS    15      2.858  -20.254   -7.239    0.000
ATOM      16   CA   ALA    16      5.501  -21.762   -4.962    0.000
ATOM      17   CA   GLY    17      2.853  -23.511   -2.872    0.000
ATOM      18   CA   ILE    18      3.262  -21.244    0.151    0.000
ATOM      19   CA   THR    19      0.552  -19.930    2.467    0.000
ATOM      20   CA   SER    20     -1.303  -16.620    2.252    0.000
ATOM      21   CA   ASP    21     -1.277  -16.162    6.024    0.000
ATOM      22   CA   PHE    22      2.504  -16.511    6.163    0.000
ATOM      23   CA   TYR    23      3.023  -14.360    3.074    0.000
ATOM      24   CA   PHE    24      0.415  -11.814    4.149    0.000
ATOM      25   CA   ASN    25      1.440  -11.988    7.803    0.000
ATOM      26   CA   ALA    26      5.136  -12.204    6.946    0.000
ATOM      27   CA   ILE    27      4.723   -9.625    4.186    0.000
ATOM      28   CA   ASN    28      2.376   -7.517    6.305    0.000
ATOM      29   CA   LYS    29      4.933   -7.369    9.113    0.000
ATOM      30   CA   ALA    30      7.715   -6.511    6.671    0.000
ATOM      31   CA   LYS    31      6.169   -3.135    5.863    0.000
ATOM      32   CA   THR    32      9.260   -1.940    4.003    0.000
ATOM      33   CA   VAL    33      9.950   -3.126    0.460    0.000
ATOM      34   CA   GLU    34     13.508   -4.173    1.292    0.000
ATOM      35   CA   GLU    35     12.244   -6.400    4.100    0.000
ATOM      36   CA   VAL    36      9.590   -7.832    1.788    0.000
ATOM      37   CA   ASN    37     12.197   -8.437   -0.909    0.000
ATOM      38   CA   ALA    38     14.497   -9.993    1.685    0.000
ATOM      39   CA   LEU    39     11.522  -11.702    3.318    0.000
ATOM      40   CA   LYS    40     10.319  -12.901   -0.081    0.000
ATOM      41   CA   ASN    41     13.838  -14.072   -0.912    0.000
ATOM      42   CA   GLU    42     14.204  -15.525    2.580    0.000
ATOM      43   CA   ILE    43     10.712  -17.015    2.429    0.000
ATOM      44   CA   LEU    44     11.242  -18.262   -1.121    0.000
ATOM      45   CA   LYS    45     14.830  -19.276   -0.388    0.000
ATOM      46   CA   ALA    46     13.882  -21.129    2.792    0.000
ATOM      47   CA   HIS    47     11.886  -23.706    0.839    0.000
```

```
1LQ7
REMARK    E=    -936.50600  Rg=    3.321  SEQ=
GSRVKALEEKVKALEEKVKALGGGGRIEELKKKWEELKKKIEELGGGGEVKKVEEEVKKLEEEIKKL
ATOM      1  CA  GLY     1       0.000   0.000   0.000   0.000
ATOM      2  CA  SER     2       3.800   0.000   0.000   0.000
ATOM      3  CA  ARG     3       3.982  -3.796   0.000   0.000
ATOM      4  CA  VAL     4       3.482  -3.977  -3.763   0.000
ATOM      5  CA  LYS     5       5.879  -1.081  -4.317   0.000
ATOM      6  CA  ALA     6       8.391  -2.771  -2.020   0.000
ATOM      7  CA  LEU     7       7.791  -6.057  -3.831   0.000
ATOM      8  CA  GLU     8       8.172  -4.295  -7.177   0.000
ATOM      9  CA  GLU     9      11.222  -2.447  -5.864   0.000
ATOM     10  CA  LYS    10      12.457  -5.655  -4.246   0.000
ATOM     11  CA  VAL    11      11.414  -7.583  -7.350   0.000
ATOM     12  CA  LYS    12      13.174  -5.009  -9.521   0.000
ATOM     13  CA  ALA    13      16.156  -5.156  -7.170   0.000
ATOM     14  CA  LEU    14      15.879  -8.944  -7.058   0.000
ATOM     15  CA  GLU    15      15.297  -9.034 -10.812   0.000
ATOM     16  CA  GLU    16      18.222  -6.663 -11.319   0.000
ATOM     17  CA  LYS    17      20.258  -8.683  -8.826   0.000
ATOM     18  CA  VAL    18      18.966 -11.902 -10.378   0.000
ATOM     19  CA  LYS    19      20.233 -10.816 -13.792   0.000
ATOM     20  CA  ALA    20      23.707 -10.298 -12.343   0.000
ATOM     21  CA  LEU    21      23.558 -13.619 -10.501   0.000
ATOM     22  CA  GLY    22      24.190 -15.603 -13.680   0.000
ATOM     23  CA  GLY    23      21.762 -16.345 -16.508   0.000
ATOM     24  CA  GLY    24      20.899 -19.841 -15.293   0.000
ATOM     25  CA  GLY    25      17.691 -21.761 -14.614   0.000
ATOM     26  CA  ARG    26      16.781 -19.728 -11.536   0.000
ATOM     27  CA  ILE    27      16.678 -16.502 -13.541   0.000
ATOM     28  CA  GLU    28      14.148 -17.973 -15.965   0.000
ATOM     29  CA  GLU    29      11.669 -18.357 -13.110   0.000
ATOM     30  CA  LEU    30      11.908 -14.621 -12.458   0.000
ATOM     31  CA  LYS    31      10.629 -13.896 -15.962   0.000
ATOM     32  CA  LYS    32       7.771 -16.341 -15.427   0.000
ATOM     33  CA  LYS    33       7.277 -14.963 -11.921   0.000
ATOM     34  CA  TRP    34       7.576 -11.445 -13.327   0.000
ATOM     35  CA  GLU    35       4.576 -12.016 -15.588   0.000
ATOM     36  CA  GLU    36       2.720 -13.836 -12.816   0.000
ATOM     37  CA  LEU    37       4.013 -11.407 -10.195   0.000
ATOM     38  CA  LYS    38       3.433  -8.463 -12.528   0.000
ATOM     39  CA  LYS    39      -0.150  -9.566 -13.155   0.000
ATOM     40  CA  LYS    40      -0.568 -10.688  -9.548   0.000
ATOM     41  CA  ILE    41       1.226  -7.600  -8.251   0.000
ATOM     42  CA  GLU    42      -1.395  -5.286  -9.741   0.000
ATOM     43  CA  GLU    43      -4.219  -7.351  -8.259   0.000
ATOM     44  CA  LEU    44      -2.274  -8.073  -5.076   0.000
ATOM     45  CA  GLY    45      -3.151  -4.705  -3.549   0.000
ATOM     46  CA  GLY    46      -3.876  -6.098  -0.089   0.000
ATOM     47  CA  GLY    47      -1.476  -6.606   2.813   0.000
ATOM     48  CA  GLY    48      -2.019 -10.356   3.106   0.000
ATOM     49  CA  GLU    49      -1.503 -11.062  -0.592   0.000
ATOM     50  CA  VAL    50       1.717  -9.044  -0.610   0.000
ATOM     51  CA  LYS    51       3.298 -11.394   1.924   0.000
```

```
ATOM     52  CA  LYS    52       3.000 -14.285  -0.524        0.000
ATOM     53  CA  VAL    53       4.909 -12.321  -3.159        0.000
ATOM     54  CA  GLU    54       7.482 -11.280  -0.563        0.000
ATOM     55  CA  GLU    55       7.896 -14.925   0.428        0.000
ATOM     56  CA  GLU    56       7.999 -15.910  -3.240        0.000
ATOM     57  CA  VAL    57      10.249 -12.947  -4.016        0.000
ATOM     58  CA  LYS    58      12.307 -13.610  -0.891        0.000
ATOM     59  CA  LYS    59      12.146 -17.351  -1.538        0.000
ATOM     60  CA  LEU    60      13.132 -16.747  -5.158        0.000
ATOM     61  CA  GLU    61      15.932 -14.430  -4.048        0.000
ATOM     62  CA  GLU    62      16.884 -16.824  -1.254        0.000
ATOM     63  CA  GLU    63      16.557 -19.803  -3.589        0.000
ATOM     64  CA  ILE    64      18.490 -18.002  -6.321        0.000
ATOM     65  CA  LYS    65      21.003 -16.597  -3.840        0.000
ATOM     66  CA  LYS    66      21.936 -20.074  -2.623        0.000
ATOM     67  CA  LEU    67      23.718 -20.892  -5.877        0.000

1CLB
REMARK    E=  -1053.83000  Rg=    2.961   SEQ=
KSPEELKGIFEKYAAKEGDPNQLSKEELKLLLQTEFPSLLKGGSTLDELFEELDKNGDGEVSFEEFQVLVKKISQ
ATOM      1  CA  LYS     1       3.800   0.000   0.000        0.000
ATOM      2  CA  SER     2       3.976  -3.796   0.000        0.000
ATOM      3  CA  PRO     3       3.124  -5.788  -3.122        0.000
ATOM      4  CA  GLU     4      -0.513  -6.749  -3.659        0.000
ATOM      5  CA  GLU     5      -0.003 -10.148  -2.039        0.000
ATOM      6  CA  LEU     6       1.461  -8.549   1.082        0.000
ATOM      7  CA  LYS     7      -1.406  -6.062   1.267        0.000
ATOM      8  CA  GLY     8      -3.999  -8.835   1.108        0.000
ATOM      9  CA  ILE     9      -2.000 -11.228   3.281        0.000
ATOM     10  CA  PHE    10      -1.421  -8.601   5.965        0.000
ATOM     11  CA  GLU    11      -5.040  -7.456   5.787        0.000
ATOM     12  CA  LYS    12      -6.284 -11.043   5.645        0.000
ATOM     13  CA  TYR    13      -3.584 -12.208   8.051        0.000
ATOM     14  CA  ALA    14      -4.184  -9.280  10.397        0.000
ATOM     15  CA  ALA    15      -7.945  -9.587   9.949        0.000
ATOM     16  CA  LYS    16      -7.904 -13.112  11.368        0.000
ATOM     17  CA  GLU    17      -6.120 -11.961  14.519        0.000
ATOM     18  CA  GLY    18      -9.401 -11.106  16.234        0.000
ATOM     19  CA  ASP    19     -11.188 -14.242  15.048        0.000
ATOM     20  CA  PRO    20      -8.781 -16.566  16.850        0.000
ATOM     21  CA  ASN    21      -5.119 -16.591  15.836        0.000
ATOM     22  CA  GLN    22      -5.813 -18.303  12.515        0.000
ATOM     23  CA  LEU    23      -2.978 -16.394  10.854        0.000
ATOM     24  CA  SER    24      -0.829 -19.521  10.635        0.000
ATOM     25  CA  LYS    25      -3.534 -21.447   8.787        0.000
ATOM     26  CA  GLU    26      -4.778 -18.348   6.974        0.000
ATOM     27  CA  GLU    27      -1.231 -17.178   6.272        0.000
ATOM     28  CA  LEU    28      -0.205 -20.714   5.332        0.000
ATOM     29  CA  LYS    29      -3.383 -21.132   3.292        0.000
ATOM     30  CA  LEU    30      -2.955 -17.641   1.851        0.000
ATOM     31  CA  LEU    31       0.777 -18.190   1.391        0.000
ATOM     32  CA  LEU    32       0.177 -21.601  -0.174        0.000
ATOM     33  CA  GLN    33      -2.302 -20.129  -2.650        0.000
ATOM     34  CA  THR    34       0.487 -18.292  -4.464        0.000
```

```
ATOM     35  CA  GLU    35       3.508 -20.250  -3.246     0.000
ATOM     36  CA  PHE    36       5.833 -18.729  -0.655     0.000
ATOM     37  CA  PRO    37       9.394 -19.681   0.271     0.000
ATOM     38  CA  SER    38       9.895 -22.379   2.899     0.000
ATOM     39  CA  LEU    39       9.260 -19.969   5.768     0.000
ATOM     40  CA  LEU    40       5.523 -20.647   5.667     0.000
ATOM     41  CA  LYS    41       5.946 -24.112   7.167     0.000
ATOM     42  CA  GLY    42       4.985 -24.665  10.802     0.000
ATOM     43  CA  GLY    43       8.547 -24.681  12.125     0.000
ATOM     44  CA  SER    44       9.508 -21.396  10.473     0.000
ATOM     45  CA  THR    45       9.413 -17.749  11.535     0.000
ATOM     46  CA  LEU    46       5.682 -17.497  10.859     0.000
ATOM     47  CA  ASP    47       4.919 -19.987  13.626     0.000
ATOM     48  CA  GLU    48       7.477 -18.340  15.903     0.000
ATOM     49  CA  LEU    49       6.361 -14.906  14.716     0.000
ATOM     50  CA  PHE    50       2.718 -15.837  15.267     0.000
ATOM     51  CA  GLU    51       3.483 -17.194  18.733     0.000
ATOM     52  CA  GLU    52       6.009 -14.437  19.411     0.000
ATOM     53  CA  LEU    53       3.816 -11.836  17.718     0.000
ATOM     54  CA  ASP    54       0.794 -12.951  19.735     0.000
ATOM     55  CA  LYS    55       2.502 -11.873  22.954     0.000
ATOM     56  CA  ASN    56       2.499  -8.232  21.868     0.000
ATOM     57  CA  GLY    57      -1.036  -7.432  20.729     0.000
ATOM     58  CA  ASP    58      -0.264  -3.878  19.625     0.000
ATOM     59  CA  GLY    59      -0.781  -2.883  15.994     0.000
ATOM     60  CA  GLU    60       2.925  -2.518  15.236     0.000
ATOM     61  CA  VAL    61       3.553  -6.177  16.045     0.000
ATOM     62  CA  SER    62       1.069  -7.290  13.394     0.000
ATOM     63  CA  PHE    63       2.493  -4.819  10.883     0.000
ATOM     64  CA  GLU    64       6.063  -5.580  11.941     0.000
ATOM     65  CA  GLU    65       5.341  -9.311  11.907     0.000
ATOM     66  CA  PHE    66       3.333  -8.942   8.702     0.000
ATOM     67  CA  GLN    67       6.046  -6.766   7.172     0.000
ATOM     68  CA  VAL    68       8.743  -8.965   8.699     0.000
ATOM     69  CA  LEU    69       6.899 -12.147   7.743     0.000
ATOM     70  CA  VAL    70       5.823 -10.668   4.412     0.000
ATOM     71  CA  LYS    71       9.274  -9.208   3.781     0.000
ATOM     72  CA  LYS    72      10.947 -12.381   5.036     0.000
ATOM     73  CA  ILE    73       8.721 -14.570   2.870     0.000
ATOM     74  CA  SER    74       9.364 -12.515  -0.261     0.000
ATOM     75  CA  GLN    75      12.880 -13.884  -0.714     0.000

1E0G
REMARK    E=    -634.31000   Rg=    2.788   SEQ=
DSITYRVRKGDSLSSIAKRHGVNIKDVMRWNSDTANLQPGDKLTLFVK
ATOM      1  CA  ASP     1       3.800   0.000   0.000     0.000
ATOM      2  CA  SER     2       4.000  -3.795   0.000     0.000
ATOM      3  CA  ILE     3       2.380  -5.979  -2.654     0.000
ATOM      4  CA  THR     4       0.381  -9.176  -2.179     0.000
ATOM      5  CA  TYR     5       0.735 -12.228  -4.416     0.000
ATOM      6  CA  ARG     6      -1.201 -15.467  -4.863     0.000
ATOM      7  CA  VAL     7       0.630 -18.763  -5.330     0.000
ATOM      8  CA  ARG     8      -2.417 -20.557  -6.722     0.000
ATOM      9  CA  LYS     9      -1.398 -19.983 -10.337     0.000
```

```
ATOM     10  CA  GLY    10       2.113 -20.872 -11.487       0.000
ATOM     11  CA  ASP    11       2.568 -17.723 -13.566       0.000
ATOM     12  CA  SER    12       2.184 -15.486 -10.519       0.000
ATOM     13  CA  LEU    13       5.898 -14.686 -10.458       0.000
ATOM     14  CA  SER    14       5.846 -13.766 -14.145       0.000
ATOM     15  CA  SER    15       2.516 -11.973 -13.779       0.000
ATOM     16  CA  ILE    16       3.619 -10.353 -10.524       0.000
ATOM     17  CA  ALA    17       7.009  -9.509 -12.020       0.000
ATOM     18  CA  LYS    18       5.327  -8.359 -15.228       0.000
ATOM     19  CA  ARG    19       2.710  -6.425 -13.267       0.000
ATOM     20  CA  HIS    20       5.382  -4.557 -11.315       0.000
ATOM     21  CA  GLY    21       8.134  -3.736 -13.802       0.000
ATOM     22  CA  VAL    22      10.701  -5.987 -12.134       0.000
ATOM     23  CA  ASN    23      12.844  -8.616 -13.847       0.000
ATOM     24  CA  ILE    24      12.715 -12.205 -12.603       0.000
ATOM     25  CA  LYS    25      15.706 -11.687 -10.317       0.000
ATOM     26  CA  ASP    26      14.094  -8.587  -8.822       0.000
ATOM     27  CA  VAL    27      10.802 -10.439  -8.409       0.000
ATOM     28  CA  MET    28      12.634 -13.536  -7.188       0.000
ATOM     29  CA  ARG    29      14.809 -11.381  -4.937       0.000
ATOM     30  CA  TRP    30      11.766  -9.328  -3.954       0.000
ATOM     31  CA  ASN    31       9.690 -12.494  -3.626       0.000
ATOM     32  CA  SER    32      12.348 -14.094  -1.432       0.000
ATOM     33  CA  ASP    33      12.345 -11.133   0.949       0.000
ATOM     34  CA  THR    34       8.569 -10.743   0.782       0.000
ATOM     35  CA  ALA    35       5.761 -12.520   2.626       0.000
ATOM     36  CA  ASN    36       3.977 -15.679   1.496       0.000
ATOM     37  CA  LEU    37       0.468 -16.459   2.729       0.000
ATOM     38  CA  GLN    38      -1.283 -19.820   2.453       0.000
ATOM     39  CA  PRO    39      -5.050 -19.821   1.958       0.000
ATOM     40  CA  GLY    40      -5.222 -23.329   0.508       0.000
ATOM     41  CA  ASP    41      -3.098 -22.329  -2.480       0.000
ATOM     42  CA  LYS    42      -0.516 -19.743  -1.442       0.000
ATOM     43  CA  LEU    43      -0.723 -15.961  -1.129       0.000
ATOM     44  CA  THR    44       2.484 -13.950  -1.462       0.000
ATOM     45  CA  LEU    45       3.125 -10.504   0.007       0.000
ATOM     46  CA  PHE    46       5.520  -8.402  -2.064       0.000
ATOM     47  CA  VAL    47       7.592  -5.229  -1.785       0.000
ATOM     48  CA  LYS    48       8.153  -3.212  -4.956       0.000

1IGD
REMARK    E=    -751.76700  Rg=     3.349  SEQ=
MTPAVTTYKLVINGKTLKGETTTKAVDAETAEKAFKQYANDNGVDGVWTYDDATKTFTVTE
ATOM      1  CA  MET     1       3.800   0.000   0.000       0.000
ATOM      2  CA  THR     2       3.907  -3.799   0.000       0.000
ATOM      3  CA  PRO     3       4.933  -5.977  -2.939       0.000
ATOM      4  CA  ALA     4       2.295  -6.653  -5.590       0.000
ATOM      5  CA  VAL     5       1.617  -9.983  -7.291       0.000
ATOM      6  CA  THR     6       0.096 -10.395 -10.749       0.000
ATOM      7  CA  THR     7      -1.317 -13.700 -11.983       0.000
ATOM      8  CA  TYR     8      -2.312 -14.533 -15.554       0.000
ATOM      9  CA  LYS     9      -4.520 -17.533 -16.308       0.000
ATOM     10  CA  LEU    10      -5.036 -19.088 -19.737       0.000
ATOM     11  CA  VAL    11      -7.874 -21.301 -18.517       0.000
```

```
ATOM     12  CA  ILE    12      -9.430 -22.386 -15.224   0.000
ATOM     13  CA  ASN    13      -8.434 -25.723 -13.704   0.000
ATOM     14  CA  GLY    14      -9.232 -27.337 -10.358   0.000
ATOM     15  CA  LYS    15      -7.590 -26.055  -7.179   0.000
ATOM     16  CA  THR    16      -4.174 -25.609  -8.781   0.000
ATOM     17  CA  LEU    17      -2.495 -24.057 -11.817   0.000
ATOM     18  CA  LYS    18      -1.890 -25.886 -15.093   0.000
ATOM     19  CA  GLY    19       0.807 -24.591 -17.435   0.000
ATOM     20  CA  GLU    20       1.729 -21.652 -15.209   0.000
ATOM     21  CA  THR    21       4.095 -18.748 -15.846   0.000
ATOM     22  CA  THR    22       5.445 -16.359 -13.216   0.000
ATOM     23  CA  THR    23       6.722 -12.785 -13.403   0.000
ATOM     24  CA  LYS    24       8.627 -11.098 -10.581   0.000
ATOM     25  CA  ALA    25       8.211  -7.612 -12.037   0.000
ATOM     26  CA  VAL    26       5.374  -6.880  -9.617   0.000
ATOM     27  CA  ASP    27       5.202 -10.565  -8.706   0.000
ATOM     28  CA  ALA    28       3.458 -11.613 -11.916   0.000
ATOM     29  CA  GLU    29       2.062 -15.085 -12.576   0.000
ATOM     30  CA  THR    30       0.826 -16.536 -15.863   0.000
ATOM     31  CA  ALA    31      -1.282 -19.694 -16.024   0.000
ATOM     32  CA  GLU    32      -2.234 -21.892 -18.974   0.000
ATOM     33  CA  LYS    33      -4.854 -23.739 -16.933   0.000
ATOM     34  CA  ALA    34      -5.560 -21.575 -13.890   0.000
ATOM     35  CA  PHE    35      -7.402 -22.485 -10.693   0.000
ATOM     36  CA  LYS    36     -11.071 -23.245 -10.058   0.000
ATOM     37  CA  GLN    37     -11.015 -21.236  -6.833   0.000
ATOM     38  CA  TYR    38     -10.117 -18.090  -8.767   0.000
ATOM     39  CA  ALA    39     -13.117 -18.450 -11.071   0.000
ATOM     40  CA  ASN    40     -15.379 -19.272  -8.131   0.000
ATOM     41  CA  ASP    41     -15.010 -15.825  -6.575   0.000
ATOM     42  CA  ASN    42     -16.214 -14.017  -9.693   0.000
ATOM     43  CA  GLY    43     -12.747 -12.774 -10.630   0.000
ATOM     44  CA  VAL    44     -13.178 -13.427 -14.349   0.000
ATOM     45  CA  ASP    45     -13.797  -9.764 -15.151   0.000
ATOM     46  CA  GLY    46     -10.770  -9.504 -17.433   0.000
ATOM     47  CA  VAL    47      -8.235  -9.251 -14.614   0.000
ATOM     48  CA  TRP    48      -8.376 -10.560 -11.049   0.000
ATOM     49  CA  THR    49      -6.792  -9.100  -7.919   0.000
ATOM     50  CA  TYR    50      -5.401 -10.759  -4.795   0.000
ATOM     51  CA  ASP    51      -4.779  -8.865  -1.560   0.000
ATOM     52  CA  ASP    52      -2.812 -11.693   0.045   0.000
ATOM     53  CA  ALA    53       0.488 -10.010  -0.806   0.000
ATOM     54  CA  THR    54      -1.279  -7.586  -3.139   0.000
ATOM     55  CA  LYS    55      -1.927 -10.046  -5.962   0.000
ATOM     56  CA  THR    56      -3.399  -9.523  -9.426   0.000
ATOM     57  CA  PHE    57      -4.823 -12.297 -11.599   0.000
ATOM     58  CA  THR    58      -5.246 -12.124 -15.371   0.000
ATOM     59  CA  VAL    59      -7.069 -14.532 -17.677   0.000
ATOM     60  CA  THR    60      -6.253 -15.266 -21.315   0.000
ATOM     61  CA  GLU    61      -8.641 -16.763 -23.864   0.000
```